\RequirePackage{fix-cm}
\documentclass{svjour3}                     % onecolumn (standard format)
\usepackage[a4paper, total={6.8in, 10in}]{geometry}
\usepackage{amsmath,epsfig}
\usepackage{multicol}
\usepackage{hyperref}
\begin{document}

\title{An Anisotropic Stellar Model
}

\author{D. Kokkinos \and  D. Pliakis \and
        T. Papakostas %etc.
}

\institute{D. Kokkinos \at		
             \email{ kokkinos@physics.uoc.gr}            \\ 
%            \emph{Present address:} of F. Author  %  if needed
\and
	D. Pliakis \at
	Department of Electronics, Technological Educational Institute of Crete\\
	Romanou 3, Chalepa, 73133, Chania, Greece\\
	\email{dpliakis@gmail.com}\\
\and
           T. Papakostas \at
              Department of Physics, University of Crete\\
	 710-03 Heraklion, Greece\\
	\email{taxiar@physics.uoc.gr}\\
}

\date{Received: date / Accepted: date}

\maketitle

\begin{abstract}

We present the matching of two solutions belonging both to Carter's family [A] of metrics. The interior solution has been found by one of us \cite{Papakostas2000} and represents an anisotropic fluid, the exterior solution is the vacuum member of Carter's family \cite{Carter1968a} of metrics. We study the model resulting from the matching procedure and we give some perspectives of our work.

\end{abstract}

%%%%%%%%%%%%%%%%%%%%%%%%%%%%%%%%%%%%%
\section{Introduction}

The alternative to the proposition statement that the Kerr solution represents a black hole, concerns the search for an interior metric that can be matched to Kerr metric, along a surface of zero hydrostatic pressure. Krasinski's paper \cite{Krasinski1978} proved to be the most complete and consistent review paper among not only new solutions, either exact or approximate, but also attempts  to match them with an appropriate vacuum solution as well  \cite{Hernandez1967}-\cite{Kyriakopoulos}.\footnote{ An alternative approach in the realm of Krasinski paper \cite{Krasinski1978} can be found in \cite{Haggag1981}, \cite{Herrera1982}.}

In the next section we begin with a short review of the Carter's family [A] of metrics and of the Complex Vectorial Formalism (CVF) of Cahen, Defrise and Debever \cite{Debever1964}, \cite{Debever1979}, \cite{Debever1981} in Newman and Penrose notation, used for our calculations. In section 3 we describe Papakostas solution and its physical interpretation. In the fourth section of this work we present the matching of Papakostas solution to the vacuum member of Carter's family [A] of metrics. In the last sections we analyse the solution and we discuss the conclusions and the impacts of the obtained results as well. 

\vspace{1.5cm}

%%%%%%%%%%%%%%%%%%%%%%%%%%%%%%%%%%%%%%%
\section{Carter's family [A] of metrics and the Complex Vectorial Formalism in Newman-Penrose notation}
\label{sec:1}

The Carter's family [A] of metrics is a class of stationary and axially symmetric spaces, characterized by the existence of a Killing tensor with two double eigenvalues or equivalently by the fact that the Hamilton-Jacobi equations for the geodesics is solvable by seperation of variables which takes place in a particular way and gives rise to a quadratic in the velocities integral of motion, see \cite{Carter1968b} and references cited there. In the CVF in the Newman-Penrose (NP) notation \cite{Debever1964}, \cite{Debever1979}, \cite{Debever1981}, the Killing tensor can be written as follows.

\begin{equation} K_{\mu \nu} = \lambda_1  (n_\mu l_\nu + l_\mu n_\nu) + \lambda_2 ( \bar{m}_\mu m_\nu + m_\mu \bar{m}_\nu ) \end{equation} 

\vspace{0.5cm}

The metric is:

\begin{equation} ds^2 = 2(\theta^1 \theta^2 - \theta^3 \theta^4) \end{equation}

\vspace{0.5cm}

where $\theta$'s are four independent Pfaffian forms in their general forms defined by the following relations. These four vectors form a covariant null orthogonal tetrad.

\begin{equation} \theta^1 = n_\mu dx^\mu \hspace{0.8cm} \theta^2 = l_\mu dx^\mu \hspace{0.8cm} \theta^3 = - \bar{m}_\mu dx^\mu \hspace{0.8cm} \theta^4 = - m_\mu dx^\mu  \end{equation}

\vspace{0.5cm}

The basis for the space ($C_3$) of complex self-dual bivector (2-forms) is given by

\begin{equation}Z^1 = \theta^1 \wedge \theta^2 \hspace{0.8cm} Z^2 = \theta^1 \wedge \theta^2 - \theta^3 \wedge \theta^4 \hspace{0.8cm} Z^3 =  \theta^4 \wedge \theta^2  \end{equation}

\vspace{0.5cm}

The component of the metric in this base are 

\begin{equation} \gamma^{a b} = 4({\delta^{a}}_{(1} {\delta^b}_{3)} -  {\delta^{a}}_{(2} {\delta^b}_{2)} ) \end{equation}

\vspace{0.5cm}

The complex connection 1-forms are defined by 

\begin{equation}dZ^a = \sigma^{a}_b \wedge Z^b \end{equation}

\vspace{2cm}

The vectorial connection 1-form is defined by 

\begin{equation}\sigma_a = \frac{1}{8} e_{a b c} \gamma^{c d} \sigma^{b}_{d}\end{equation}

\vspace{0.5cm}

Where $ e_{a b c} $ is the Levi-Civita tensor. 
The connection coefficients $\sigma_{a} = \sigma_{a \alpha}  \theta^{\alpha}$ can be expressed in terms of the 12 complex spin coefficients.

\begin{equation}  \sigma_{a \alpha} = \begin{pmatrix}  \kappa & \tau & \sigma & \rho \\ \epsilon & \gamma & \beta & \alpha \\ \pi & \nu & \mu & \lambda\\          \end{pmatrix} \end{equation}

\vspace{0.2cm}

The complex curvature 2-forms $ \Sigma^{b}_{d} $ are defined by 

\begin{equation} \Sigma^{b}_{d} = d \sigma_{d}^{b} + \sigma^{b}_{g} \wedge \sigma^{g}_{d} \end{equation}

\vspace{0.5cm}

and the vectorial curvature 2-form by 

\begin{equation}\Sigma_{a} = \frac{1}{8} e_{a b g } \gamma^{b g } \end{equation}

\vspace{0.5cm}

On expanding $\Sigma_a$ in the basis of $ \left[ Z^a , \bar{Z}^a \right]$ we can obtain

\begin{equation}\Sigma_a = (C_{a b} - \frac{1}{6} R \gamma_{ a b})Z^b + E_{a b} \bar{Z}^b \end{equation}

\vspace{0.5cm}

where these tensors are related with the curvature components $ \Psi_{A}$  and $\Phi_{A B} $ of the formalism

\begin{equation}C_{ a b} = \begin{pmatrix}  \Psi_0 & \Psi_1 & \Psi_2 \\ \Psi_1 & \Psi_2 &\Psi_3 \\ \Psi_2 & \Psi_3 & \Psi_4\\          \end{pmatrix},  \hspace{1cm} E_{ a b} = \begin{pmatrix}  \Phi_{00} & \Phi_{01} & \Phi_{02} \\ \Phi_{10} & \Phi_{11} &\Phi_{12} \\ \Phi_{20} & \Phi_{21} & \Phi_{22}\\          \end{pmatrix}\end{equation}
\vspace{1cm}

The Carter's family of metrics splits to four subfamilies according to the status of the eigenvalues $\lambda_1$ and $\lambda_2$. If $\lambda_1$ and $\lambda_2$ are not constants we have the family [A], if $\lambda_1$ is constant and $\lambda_2$ is not, we have the family [B(+)], if $\lambda_2$ is constant and $\lambda_1$ is not, we have the family [B(-)], and finally if both eigenvalues are constants we have the family [D]. In this paper we will study the class [A] of Carter's metrics exclusively, which can be written as follows:

\small
\begin{equation}ds^2 =   \frac{f  E^2(y)}{(x^2 + y^2)}(dt -x^2dz)^2 - \frac{H^2 (x)}{(x^2+y^2)}(dt +y^2dz)^2 -(x^2 + y^2) \left[ \frac{y^2 dy^2}{G^2 (y)} + \frac{x^2 dx^2}{F^2 (x)}   \right] \end{equation}
\normalsize

In this paper we consider only the case where $f=\pm 1$, which denotes the existence of one time-like Killing vector and one space-like Killing vector $\frac{\partial}{\partial t}$ and $\frac{\partial}{\partial z}$ respectively, associated with the stationarity and axial symmetry of the metric \cite{Papakostas1983}.

The vacuum member of this class of metrics is given by the following setting:

$$G^2(y) = y^2 E^2(y), \hspace{2cm} F^2(x) = x^2 H^2 (x)$$
\begin{equation} \end{equation}
$$ E^2(y) = \frac{b}{2}y^2 +dy + p, \hspace{3cm} H^2(x) = - \frac{b}{2}x^2 + cx+p$$

With this defining of $E^2(y), H^2(x)$ we have the Kerr metric with or without  Cosmological Constant.  

\vspace{0.2cm}
With the appropriate choice of our constants and with the transformation of our coordinates.
$$  b=2, \hspace{0.5cm} d=-2m \hspace{0.5cm} c = 0  \hspace{0.5cm} p = \alpha^2 $$
 
$$ y = r, \hspace{1cm} x = \alpha cos(\theta), \hspace{1cm} z= \frac{\phi}{\alpha}, \hspace{1cm} t = \tilde{t} + \alpha \phi$$

\vspace{0.5cm}

and       

$$\tilde{t} \to - t  $$

where m is the mass, $\alpha$ is the angular momentum per unit mass, c is the NUT parameter, the presence of which denotes that the spaces are not asymptotically flat. Hence, we take the vacuum member of Carter's family [A] of metrics.

\vspace{0.5cm}

\small
$$ds^2 =  \frac{\Delta - \alpha^2 sin^2 \theta}{\rho^2} dt^2 + 2 \alpha \frac{2mrsin^2 \theta}{\rho^2}dt d\phi - \frac{\rho^2}{\Delta}dr^2 $$
\normalsize
\begin{equation}    \end{equation}
\small
$$- \rho^2 d\theta^2 - \frac{(r^2+\alpha^2)^2 - \alpha^2 \Delta sin^2\theta}{\rho^2}sin^2\theta d\phi^2$$
   
\normalsize

\vspace{0.9cm}

\begin{equation}where \hspace{1 cm} \Delta \equiv r^2 - 2mr+ \alpha^2, \hspace{1.5cm} {\rho}^2 \equiv r^2 + \alpha^2 cos^2\theta \end{equation}

\vspace{1.5cm}

%%%%%%%%%%%%%%%%%%%%%%%%%%%%%%%%%%%%%%%%%%
\section{The Papakostas Solution}
\label{sec:2}

The components of the traceless Ricci tensor and the Weyl tensor in the NP notation for the metric (13) are the following relations

\begin{equation}\Phi_{01} = \Phi_{21} = \frac{H(x) E(y)}{4 ( x^2 + y^2)^3}  \left[ \frac{{T^{2}}_y }{2y} \hspace{0.1cm}  -  \hspace{0.1cm} \frac{{\Pi^2}_x}{2x}  \right]\end{equation}
\vspace{0.2cm}

\begin{equation}\Phi_{00} = \Phi_{22} = \frac{ E^2 (y)}{2 ( x^2 + y^2)^3}  \left[ \frac{{T^{2}} + {\Pi}^2 }{x^2+y^2}  - \frac{{T^2}_y}{2y} \right] \end{equation}

\vspace{0.2cm}

\begin{equation}\Phi_{02} = \Phi_{20} = \frac{ H^2 (y)}{2 ( x^2 + y^2)^3}  \left[ - \frac{{T^{2}} + {\Pi}^2 }{x^2+y^2}  + \frac{{\Pi^2}_x}{2x} \right] \end{equation}

\vspace{0.2cm}

\begin{equation} \Psi_1 = \Psi_3 =  \frac{H(x)E(y)}{2(x^2+y^2)^2 } \left[ - \frac{2(T^2 + \Pi^2)}{x^2+y^2} + \frac{(T^2)_y}{2y} + \frac{(\Pi^2)_x}{2x}  \right] \end{equation}

\vspace{0.2cm}

\begin{equation} \Psi_0 = \Psi_4 = 0 \end{equation}

\vspace{0.2cm} 

$\Phi_{11}$, $6\Lambda$, $\Psi_2$ are lengthly and they will not be given at this moment. Also we used the notation

$$T(y) \equiv \frac{G(y)}{E(y)} \hspace{2cm} \Pi (x) \equiv \frac{F(x)}{H(x)}$$

\begin{equation} \end{equation}

$$ {T^{2}}_y   = \frac{dT^2}{dy^2}     \hspace{2cm}   {\Pi^{2}}_x = \frac{d \Pi^2}{dx^2}$$

\vspace{1cm}

We assume that the energy momentum tensor is locally anisotropic and it possesses four distinct eigenvalues. In the tangent space the energy-momentum tensor can be put in the form

\begin{equation} {T^{\mu}}_{\nu} =  \begin{bmatrix} e(x,y) & 0 & 0 & 0\\ 0 & p_{y}(x,y) & 0 & 0\\  0 & 0 &  p_{z}(x,y)  & 0\\  0 & 0 & 0 &  p_{x}(x,y)\\           \end{bmatrix}   \end{equation}

\vspace{0.8cm}

In the tetrad defined by (3) the energy momentum tensor is given by:

$$ T_{\mu \nu} = \frac{1}{2}(e + p_y)(n_\mu n_\nu + l_\mu l_\nu) +  \frac{1}{2}( p_z - p_x)(\bar{m}_\mu \bar{m}_\nu + m_\mu m_\nu)  $$

\begin{equation} +   \frac{1}{4}(e - p_y + p_z + p_x)(n_\mu l_\nu+l_\mu n_\nu + \bar{m}_\mu m_\nu + m_\mu \bar{m}_\nu ) 	 \end{equation}

$$ +  \frac{1}{4}(e-p_y-p_z-p_x)( n_\mu l_\nu+ l_\mu n_\nu - \bar{m}_\mu m_\nu -m_\mu \bar{m}_\nu        )   $$

\vspace{1cm}

The Einstein's equations 

$$R_{\mu \nu} - \frac{R}{2} g_{\mu \nu} = T_{\mu \nu}$$

must be written in CVF in the NP notation in order to permit us to define the components of the traceless Ricci tensor.

\begin{equation}\Phi_{01} = 0 = \Phi_{12}    \end{equation}
\begin{equation}\Phi_{00}=\frac{e+ p_y}{4}\end{equation}
\begin{equation}\Phi_{02} = \frac{p_z - p_x}{4}\end{equation} 
\begin{equation} 2\Phi_{11} = \frac{e-p_y+p_z+p_x}{4}\end{equation} 
\begin{equation}\frac{R}{4} = - \frac{e-p_y - p_z - p_x}{4}= -6 \Lambda\end{equation}

The equivalence between $\Phi's$ with the four eigenvalues gives the form of the eigenvalues\footnote{The first eigenvalue represents the mass-energy density, the other three eigenvalues represent  the stresses.}.

\begin{equation} e(x,y)  = 6\Lambda + 5 \Phi_{00} + 3 \Phi_{02}  \end{equation}
\begin{equation} p_x(x,y) =  - 6 \Lambda +3 \Phi_{00} +   \Phi_{02}  \end{equation}
\begin{equation} p_y(x,y) =  - 6 \Lambda -\Phi_{00} - 3\Phi_{02}  \end{equation}
\begin{equation} p_z(x,y) =  - 6 \Lambda + 3 \Phi_{00} +5\Phi_{02}   \end{equation}

\vspace{0.7cm}

The equation (25) implies that 

\begin{equation} \frac{(T^2)_y}{2y} -  \frac{(\Pi^2)_x}{2x} =0 \end{equation}

This equation is easily solved as we can see below.

$$ T^2 (y) = \frac{G^2 (y)}{E^2 (y)}= k_2 y^2 + l_0 $$ 
\begin{equation} \end{equation}
$$ \Pi^2 (x) = \frac{F^2 (x)}{H^2 (x)}= k_2 x^2 + l_0 $$

We can now write the expressions of $\Phi_{AB}$ and $\Psi_A$

\vspace{0.5cm}

\begin{equation}\Phi_{01} = \Phi_{21} = 0 \end{equation}

\begin{equation}\Phi_{00} = \Phi_{22} =  \frac{E^2 (y) (k_0 + l_0)}{2(x^2+y^2)^3} \end{equation}

\begin{equation}\Phi_{02} = \Phi_{20} = - \frac{H^2 (x) (k_0 + l_0)}{2(x^2+y^2)^3} \end{equation}

\begin{equation} \Psi_1 = \Psi_3 = -  \frac{H(x)E(y) (k_0 + l_0)}{2(x^2+y^2)^3 } \end{equation}

\scriptsize

$$ 2 \Phi_{11} = \frac{k_2 y^2 + k_0}{4y^2 (x^2+y^2)}E^2_{yy} - \frac{4y^2(k_2 y^2 + k_0)+k_0(x^2+y^2)}{4 y^3 (x^2+y^2)^2}E^2_y + \left[ \frac{2 k_2(x^2 + y^2)+3(k_0 + l_0)}{2(x^2+y^2)^3}   \right] E^2  $$

\begin{equation}- \frac{k_2 x^2 + l_0}{4x^2 (x^2+y^2)}H^2_{xx} + \frac{4x^2(k_2 x^2 + l_0)+l_0(x^2+y^2)}{4 x^3 (x^2+y^2)^2}H^2_x - \left[ \frac{2k_2(x^2 + y^2) + 3 (k_0 + l_0)}{2(x^2+y^2)^3}  \right] H^2 \end{equation}

\normalsize
\vspace{0.2cm}

$$ - 6 \Lambda = \frac{k_2 y^2 + k_0}{4y^2 (x^2+y^2)}E^2_{yy} - \frac{k_0}{4 y^3 (x^2+y^2)}E^2_y -  \frac{k_0 + l_0}{2(x^2+y^2)^3}  E^2 $$

\begin{equation}- \frac{k_2 x^2 + l_0}{4x^2 (x^2+y^2)}H^2_{xx} - \frac{l_0}{4 x^3 (x^2+y^2)}H^2_x -  \frac{k_0 + l_0}{2(x^2+y^2)^3}  H^2 \end{equation}

\vspace{0.8cm}

The expression for $\Psi_2$ is lengthly and it is not given here. Two cases appear. The first case concerns the annihilation of the sum of the contants $k_0$, $l_0$.

$$    k_0 + l_0 = 0 $$

(I)
$$\Psi_1 = \Psi_3 = 0, \hspace{0.3cm} \Psi_2 \neq 0$$

\vspace{0.3cm}

This metric is of type D in the Petrov Classification and it is the Carter's family [A] of spaces. The second case concerns the non-annihilation of the previous sum.

\vspace{0.4cm}

(II) \hspace{6.5cm}   $ k_0 + l_0 \neq 0 $

\vspace{0.3cm}

Hence, in this case  the metric is of type I in the Petrov classification provided that $4{\Psi_2}^2 \neq 16{\Psi_1}^2$, and it represents an interior anisotropic solution because of the choice of the energy-momentum tensor(23).

The functions $E^2 (y)$ and $H^2 (x)$ remain undefined, for this reason we have to impose an "equation of state", a relation between the four eigenvalues. This supplementary condition has been chosen due to its convenience in the integration. This choice allow us to obtain a differential equation for the unknown functions $E^2 (y)$ and $H^2 (x)$. The relation is

\begin{equation} e + p_z = 2(p_x - p_y) \end{equation}

and can be written as follows using (25)-(28)

\begin{equation} 2 \Phi_{11} - 3 \Phi_{00} -3 \Phi_{02} = 0\end{equation}

\vspace{0.5cm}
Now, if we use the expression for $\Phi_{AB}$ (17)-(19) we take the following differential equation
\vspace{0.5cm}

\scriptsize
$$ (x^2 + y^2)(k_2 y^2 + k_0) \frac{{E^2}_{yy}}{y^2} - (4y^2( k_2 y^2 + k_0) +k_0(x^2+y^2) ) \frac{(E^2)_y}{y^3} + 4k_2 E^2$$
\begin{equation} \end{equation}
$$ -  (x^2 + y^2)(k_2 x^2 + l_0) \frac{{H^2}_{xx}}{x^2} + (4x^2( k_2 y^2 + l_0) +l_0(x^2+y^2) ) \frac{(H^2)_x}{x^3} - 4k_2 H^2 = 0$$
  \normalsize
\vspace{0.7cm}

This equation can be solved by seperation of variables, we differentiate it twice with respect to x and twice with respect to y:

\begin{equation}\left[    (k_2 y^2 + k_0) \frac{{E^2}_{yy}}{y^2} - k_0  \frac{{E^2}_{y}}{y^3}     \right]_{yy} - \left[    (k_2 x^2 + l_0) \frac{{H^2}_{xx}}{x^2} - l_0  \frac{{H^2}_{x}}{x^3}     \right]_{xx} = 0  \end{equation}

The equation splits into two equations:

\begin{equation} \left[    (k_2 y^2 + k_0) \frac{{(E^2)}_{yy}}{y^2} - k_0  \frac{{(E^2)}_{y}}{y^3}     \right]_{yy}  = 8d  \end{equation}

\begin{equation} \left[    (k_2 x^2 + l_0) \frac{{(H^2)}_{xx}}{x^2} - l_0  \frac{{(H^2)}_{x}}{x^3}     \right]_{xx} = 8d \end{equation}

\vspace{1cm}
Finally the integration of these equations gives us the relations for $E^2(y)$ and $H^2(x)$.

\begin{equation} E^2 (y) = \frac{d}{3k_2}y^4 + \left( \frac{d_2}{k_2} - \frac{4k_0 d}{3 {k_2}^2} \right) y^2 + \frac{2 k_0 d_2}{{k_2}^2} - \frac{d_0}{k_2} - \frac{8 {k_0}^2 d}{3 {k_2}^3} + q_1 \sqrt{k_2 y^2 + k_0} \end{equation}

\begin{equation} H^2 (x) = \frac{d}{3k_2}x^4 - \left( \frac{d_2}{k_2} + \frac{4l_0 d}{3 {k_2}^2} \right) x^2 - \frac{2 l_0 d_2}{{k_2}^2} - \frac{d_0}{k_2} - \frac{8 {l_0}^2 d}{3 {k_2}^3} + q_2 \sqrt{k_2 x^2 + l_0} \end{equation}

Then we can express the eigenvalues, $\Phi_{AB}$, $\Psi_A$, as functions of $E^2 (y)$ and $H^2 (x)$:

\begin{equation} 2 \Phi_{11} = \frac{3}{2} \frac{k_0 + l_0}{(x^2 + y^2)^3} \left[ E^2 (y) - H^2 (x) \right]  \end{equation}

\begin{equation} -6\Lambda = d -   \frac{1}{2} \frac{k_0 + l_0}{(x^2 + y^2)^3} \left[ E^2 (y) + H^2 (x) \right]  \end{equation}

\begin{equation}  \Phi_{00} = \Phi_{22} =  \frac{k_0 + l_0}{2(x^2 + y^2)^3}  E^2 (y)  \end{equation}

\begin{equation}  \Phi_{02} = \Phi_{20} = - \frac{k_0 + l_0}{2(x^2 + y^2)^3}  H^2 (x)  \end{equation}

\begin{equation}  \Psi_1 = \Psi_3 = - \frac{k_0 + l_0}{2(x^2 + y^2)^3}E(y)  H (x)  \end{equation}

\begin{equation} e = -d  + \frac{k_0 + l_0}{(x^2 + y^2)^3} \left[3 E^2 (y) - H^2 (x) \right]  \end{equation}

\begin{equation} p_x= d  + \frac{k_0 + l_0}{(x^2 + y^2)^3} \left[ E^2 (y) - H^2 (x) \right]  \end{equation}

\begin{equation} p_y= d  + \frac{k_0 + l_0}{(x^2 + y^2)^3} \left[- E^2 (y) - H^2 (x) \right]  \end{equation}

\begin{equation} p_z= d  + \frac{k_0 + l_0}{(x^2 + y^2)^3} \left[ E^2 (y) - 3H^2 (x) \right]  \end{equation}

The constant d is the cosmological constant, also $d_2$, $d_0$, $d$, $k_2$, $k_0$, $l_0$, $q_1$, $q_2$ are contants due to integration. The relations of $E^2 (y)$, $H^2 (x)$ gives us the first exact solution that can be interpreted as a anisotropic fluid solution, in the case of stationary and axial symmetry. Wahlquist in ref. \cite{Wahlquist1968} mentioned that the existence of these kind of solutions is possible. Also this solution reduces to the vacuum family [$\tilde{A}$] of Carter's spaces if we impose the following condition:

\begin{equation} k_0 + l_0 = 0 \end{equation}

The solution is compatible with the existence of a rotation axis which satisfies the condition of elementary flatness. The position of the rotation axis is given by the vanishing of the axial Killing vector $x=0$.

\begin{equation} H^2 (x=0) = 0 \end{equation}

This condition for our solution (49) is the following.

\begin{equation} q_2 \sqrt{l_0} + 2\frac{{l_0} d_2}{{k_2}^2} - \frac{d_0}{k_2} - \frac{8 l_0 d}{3 {k_2}^3} = 0   \end{equation}

The last condition ensures also the elementary flatness of the rotation axis 

\begin{equation}\frac{X_{,i}X^{,i} }{4X} \to 1 \end{equation}

where

\begin{equation}X = U_i U^i, \hspace{1cm} U = \frac{\partial}{\partial z}\end{equation}

%%%%%%%%%%%%%%%%%%%%%%%%%%%%%%%%%%
\section{The Soldering of the solutions}
\label{sec:4}

The "Soldering" of the solutions referred to the maching of the Papakostas solution with the vacuum member of Carter's family of metrics which both are members of the Carter's family of the solutions. The treatment of this problem will  construct a realistic stellar model. In the past, there were various works which aimed to find a possible source for known exterior gravitational fields before since the genesis of a realistic stellar model is a intriguing problem.

The matching conditions of the "Soldering", which are widely known as Darmois junction conditions \cite{Darmois1927}, ensure the success of the matching. We present the metric of Papakostas

\vspace{1cm}

\footnotesize

\begin{equation}ds^2 =   \frac{ E^2(y)}{(x^2 + y^2)}(dt -\frac{x^2}{\alpha}d\phi)^2 - \frac{H^2 (x)}{(x^2+y^2)}(dt +\frac{y^2}{\alpha}d\phi)^2 -(x^2 + y^2) \left[ \frac{y^2 dy^2}{(k_2 y^2+k_0)E^2 (y)} + \frac{x^2 dx^2}{(k_2 x^2 + l_0)H^2 (x)}   \right] \end{equation}

\normalsize 

\vspace{0.4cm}

using the following functions

\vspace{0.4cm}

\begin{equation} E^2 (y) = \frac{d}{3k_2}y^4 + \left( \frac{d_2}{k_2} - \frac{4k_0 d}{3 {k_2}^2} \right) y^2 + \frac{2 k_0 d_2}{{k_2}^2} - \frac{d_0}{k_2} - \frac{8 {k_0}^2 d}{3 {k_2}^3} + q_1 \sqrt{k_2 y^2 + k_0} \end{equation}

\begin{equation} H^2 (x) = \frac{d}{3k_2}x^4 - \left( \frac{d_2}{k_2} + \frac{4l_0 d}{3 {k_2}^2} \right) x^2 - \frac{2 l_0 d_2}{{k_2}^2} - \frac{d_0}{k_2} - \frac{8 {l_0}^2 d}{3 {k_2}^3} + q_2 \sqrt{k_2 x^2 + l_0} \end{equation}

\vspace{0.4cm}

 and the Carter's Family of metrics $[\tilde{A}]$.

\vspace{0.4cm}

\begin{equation}ds^2 =   \frac{ {E_C}^2(y)}{(x^2 + y^2)}(dt -\frac{x^2}{\alpha}d\phi)^2 - \frac{{H_C}^2 (x)}{(x^2+y^2)}(dt +\frac{y^2}{\alpha}d\phi)^2 -(x^2 + y^2) \left[ \frac{dy^2}{{E_C}^2 (y)} + \frac{dx^2}{{H_C}^2 (x)}   \right] \end{equation}

\vspace{0.4cm}

The corresponding functions are given by
\vspace{0.4cm}

\begin{equation}{E_C}^2(y) = \frac{a}{2}y^2 + \tilde{d} y + c\end{equation}

\begin{equation}{H_C}^2(x) = - \frac{a}{2}x^2 + bx + c\end{equation}

\vspace{0.3cm}

\subsection{The matching of the components}
  
The matching requirements concern mostly the components of the metrics. We require continuity of each component of the metric. In other words, we require not only the equivalence of the components of the two metrics but also the equivalence of their first derivatives upon the surface of our star, the zero-pressure surface. Then, we will be able to compute its radius.

\small

\begin{multicols}{2}

\begin{center}
Interior

Anisotropic Fluid Solution

$$\frac{E^2(y) - H^2(x)}{x^2+y^2}$$

$$-\frac{x^2 E^2(y) + y^2 H^2(x)}{\alpha(x^2+y^2)}$$

$$\frac{x^4 E^2(y) - y^4 H^2(x)}{\alpha^2(x^2+y^2)}$$

$$y^2\frac{(x^2+y^2)}{(k_2 y^2 + k_0) E^2(y)}$$

$$x^2\frac{(x^2+y^2)}{(k_2 x^2 + l_0) H^2(x)}$$

\end{center}

\columnbreak
\begin{center}

Exterior

Carter's Metric

\begin{equation}\frac{{E_C}^2(y) - {H_C}^2(x)}{x^2+y^2}\end{equation}

\begin{equation}- \frac{x^2{E_C}^2(y) + y^2{H_C}^2(x)}{\alpha(x^2+y^2)}\end{equation}

\begin{equation} \frac{x^4{E_C}^2(y) - y^4{H_C}^2(x)}{\alpha^2(x^2+y^2)}\end{equation}

\begin{equation}y^2\frac{(x^2+y^2)}{y^2 {E_C}^2(y)}\end{equation}

\begin{equation}x^2\frac{(x^2+y^2)}{x^2 {H_C}^2(x)}\end{equation}

\end{center}
\end{multicols}

\vspace{2cm}
\normalsize

These relations represent the components of the metrics and correspond accordingly to $g_{tt}, g_{t\phi}$, $g_{\phi \phi}, g_{yy},$and $g_{xx}$. The resolving of the system of the equations demands, as a necessary procedure, the exact determination of the form of the $y$. One possible case for $y$ component is to be equal to a constant, which is the radius of our star, denoting  the shape of our zero-pressure surface is an ellipsoid of revolution (oblate spheroid)  or a toroidal since it belongs to Carter's Family of solutions. \cite{Papakostas2015}.

For these reasons we want the exact matching of the function $H^2(x)$ in order to eliminate it from every equation. The eliminating of the function $H^2(x)$ lets us obtain only two equations. The relation $(73)$ concerns the equivalence of $G^2(y)$ while the rest give the equivalence of $E^2(y)$ with ${E_C}^2(y)$ .

  The matching of the function of $x$ requires the following setting as a first step.

$$l_0 = 0 \hspace{2cm} k_2 = 1$$

The second step defines the constants of $H^2(x)$. 

$$d=0 \hspace{1cm} d_2 = \frac{a}{2} \hspace{1cm} q_2 = b \hspace{1cm} d_0 = -c$$

Due to the equivalence of $H^2(x)$ with ${H^2}_C (x)$, any of these functions are eliminated from the previous equations. The settings above have a significant affection into ${E^2} (y)$. 

\begin{equation} E^2(y) = \frac{a}{2}y^2 + q_1 \sqrt{y^2 + k_0} + c + a k_0 \end{equation}

\begin{equation} H^2(x) = - \frac{a}{2}x^2 + b x + c \end{equation}

The next stage requires the equivalence of the function $G^2(y)$ with $G{^2}_C (y)$. One possible choice is the elimination of the constant $k_0$. Indeed, this choice achieves the equivalence of these two functions along with the elimination of the fluid which isn't a desirable result since the simultaneous elimination of the constants $k_0$ and $l_0$ gives us absence of fluid.

Now, the relations $(70)-(72)$ give

\begin{equation}E^2(y) - {E_C}^2(y) = 0 \hspace{0.5cm}  \to \hspace{0.5cm} q_1 \sqrt{y^2+k_0} + a k_0 = \tilde{d} y  \end{equation}

While from the equivalence of $G's$ relation $ (7.10)$ we take 

 \small

$$( y^2+ k_0)E^2(y) - (y^2){E_C}^2(y) = 0  \hspace{0.5cm} \to  \hspace{0.5cm}$$

\begin{equation}  (y^2 + k_0) \left[ q_1 \sqrt{y^2+k_0} + ak_0 \right] + k_o \left[ \frac{a}{2}y^2 + c \right]  = \tilde{d}y^3\end{equation}

\normalsize

\vspace{0.5cm}

\subsection{Matching of the derivatives of the components}

\vspace{1cm}

The corresponding calculations for the first derivatives of the metrics' components only in respect to $y$ according to Darmois junction conditions.

\begin{multicols}{2}

\begin{center}
Interior

Anisotropic Fluid Solution

\small

$$\frac{(E^2(y) )_y}{x^2+y^2} -2y\frac{E^2(y) - H^2(x)}{(x^2+y^2)^2}$$

$$-\frac{x^2 (E^2(y))_y + 2y H^2(x)}{\alpha((x^2+y^2)} + 2y\frac{x^2 E^2(y) + y^2 H^2(x)}{(x^2+y^2)^2}$$

$$\frac{x^4 (E^2(y))_y - 4y^3 H^2(x)}{\alpha^2(x^2+y^2)} - 2y\frac{x^4 E^2(y) - y^4 H^2(x)}{\alpha^2(x^2+y^2)^2} $$

$$\frac{2y(x^2+2y^2)}{(k_2 y^2 + k_0) E^2(y)} - \frac{y^2(x^2+y^2) \left[ 2y k_2E^2(y) + (k_2y^2+k_0) (E^2(y))_y  \right] }   {(k_2 y^2 + k_0)^2 E^4(y)}$$

$$\frac{2yx^2}{(k_2 x^2 + l_0) H^2(x)}$$

\end{center}

\columnbreak
\begin{center}

Exterior

Carter's Metric

\small

\begin{equation}\frac{({E_C}^2(y) )_y}{x^2+y^2} -2y\frac{{E_C}^2(y) - {H_C}^2(x)}{(x^2+y^2)^2}\end{equation}

\begin{equation}-\frac{x^2 ({E}_C^2(y))_y + 2y {H_C}^2(x)}{\alpha((x^2+y^2)}- 2y\frac{x^2{E_C}^2(y) + y^2{H_C}^2(x)}{\alpha(x^2+y^2)^2}\end{equation}

\begin{equation} \frac{x^4 ({E_C}^2(y))_y - 4y^3{H_C}^2(x)}{\alpha^2(x^2+y^2)} - 2y \frac{x^4{E_C}^2(y) - y^4{H_C}^2(x)}{\alpha^2(x^2+y^2)^2}\end{equation}

\begin{equation}\frac{2y(x^2+2y^2)}{y^2 {E_C}^2(y)} - \frac{y^2(x^2+y^2)  \left[ 2y {E_C}^2(y) + y^2 ({E_C}^2(y))_y  \right]   }{y^4 {E_C}^4(y)}\end{equation}

\begin{equation}\frac{2yx^2}{x^2 {H_C}^2(x)}\end{equation}

\end{center}
\end{multicols}

\normalsize

\vspace{1cm}

In these calculations the terms that don't contain the derivatives of the functions $E^2(y),{E_C}^2(y)$ are eliminated since we have achieved the equivalence between $E^2(y) , H^2(x)$ with  ${E_C}^2(y) , {H_C}^2(x)$ accordingly, already in the previous section. Resulting, our equations $(79) - (83)$ reduce to $(84)$, $(85)$. 

\begin{equation}     (E^2(y))_y -({E}_C^2(y))_y = 0  \hspace{0.3cm}  \to      \hspace{0.3cm}    \tilde{d} \sqrt{y^2 + k_0} = q_1 y           \end{equation}

\begin{equation}    (k_2y^2+k_0)( E^2(y))_y -y^2({E}_C^2(y))_y = 0  \hspace{0.3cm} \to  \hspace{0.3cm}     q_1\sqrt{y^2 + k_0}  + a k_0  =   \tilde{d}y  \end{equation}

\vspace{0.3cm}

Finally the relations (85) and (77) will be equivalent if we will use as $E^2(y)$ the relation (75).

\vspace{0.3cm}

\subsection{Constraints}

The matching of the components gives us a solution for radius $y$ and constraints for our constants. Also the relation (84) is the derivative of relation (77) with respect to $y$

\vspace{0.2cm}

$$ q_1 \sqrt{y^2+k_0} + a k_0 = \tilde{d}y \hspace{7.1cm} (77)$$

$$ (y^2 + k_0) \left[ q_1 \sqrt{y^2+k_0} + ak_0 \right] + k_o \left[ \frac{a}{2}y^2 + c \right]  = \tilde{d}y^3     \hspace{3cm} (78)$$

$$  q_1 =  \frac{\tilde{d} \sqrt{y^2 + k_0}}{y}     \hspace{8.1cm} (84) $$

\vspace{1cm}

If we substitute the value of $q_1$ to the relation (77), we will obtain the radius $y$ to be equal to a phenomenical negative value. In this point, we demand that the ratio $\tilde{d}/a$ to be a negative value. This is a physical acceptable choice since the coordinate $y$ represents the radius $r$ of the star in Boyer-Lindquist coordinates.

\begin{equation} y_h = -\frac{\tilde{d}}{a} \end{equation}

Now, if we impose again the value of $q_1$ and the relation (86) to the relation (78) , we will finally determine the unknown contants of our solution.

\begin{equation} c = \frac{\tilde{d}^2}{2a} \end{equation}

\begin{equation} q_1 = - \sqrt{\tilde{d}^2 + a^2 k_0} \end{equation}

The equations of constraints are satisfied and with this procedure our unknown constants have been determined. Finally we achieved to compute all the unknown constants of our function $E^2 (y)$. The new forms of our function and of Carter's at the point $y_h$, are the following.

\begin{equation} E^2 (y_h) = \frac{a}{2}{y_h}^2 - \sqrt{ \tilde{d}^2 + a^2 k_0} \sqrt{{y_h}^2 + k_0}   +  \frac{\tilde{d}^2}{2a} +  a k_0 \end{equation}

\begin{equation}  {E^2}_C (y_h) = \frac{a}{2}{y_h}^2 + \tilde{d}{y_h} + \frac{\tilde{d}^2}{2a} \end{equation}

These two functions match at the radius $y_h =- \tilde{d}/a$ and simultaneously are annihilated at the same point. This elimination of the functions sets $g_{rr}$ component equal to infinity. The metrics match at an \textbf{event horizon}. 

At the first sight, we see an interior anisotropic fluid solution to be contained at a generalization of Kerr-NUT solution and the matching takes place at an event horizon. Trying to find out the nature of this event horizon we see that the function ${E^2}_C (y)$ has a double root at the $y_h$, while the function $E^2 (y)$ has two roots which the plus root is negative and simultaneously consists an  unacceptable choice for the radius.

$$y_{\pm} = \pm \frac{\tilde{d}}{a}$$

If we try to take the Kerr (-NUT)\footnote{The constant which distinguishes the Kerr - NUT from Kerr concerns only the NUT constant $b$ at the function $H^2 (x)$. }  the constants of the function ${E^2}_C$ takes the following values.

$$\tilde{d} \to -2m$$

$$ a \to 2$$

$$ \frac{\tilde{d}^2}{2a} \to  m^2 $$

This results to the equivalence between the term $\alpha^2$ with the term $m^2$. This result reveals that the only case in which, we can obtain a Kerr-like as an exterior solution of our interior rotating fluid solution, is the maximum limit of Kerr which is known as \textbf{Maximal Kerr} since the Kerr constant $\alpha$ is equal to mass m.

$${ E^2}_C (y_h) = {y_h }^2 - 2my_h + m^2 $$

Concluding, the "soldering" takes place at the unique event horizon that exists in case of the Maximally rotating case of Kerr at the radius $y_h = m$. Also the inner part of the object is filled with matter characterized by the anisotropic fluid solution of Papakostas.

\section{Analysis of the solution }
\label{sec:5}

\vspace{0.7cm}

In this chapter, we try to define the pressure of the fluid intending to analyse our solution at the main regions of the star. The center of the star is one of the most interesting domains in order to decide if there are any singularities or anomalies and the surface of the star is going to enlight us about the radius and the shape of our ellipsoid.     

The strategy we will follow, in order to define the zero-pressure surface, aims to the determination of the pressure at the first place and its elimination on the surface. This computation requires to consider our diagonalized general anisotropic matrix $(23)$ as a sum of two diagonalized matrices. The first one is the perfect-fluid tensor, while  the other one contains the anisotropic stresses.
\vspace{0.7cm}

$$ T^{\mu \nu} = {T_p}^{\mu \nu} + {\Pi_A}^{\mu \nu}$$

\scriptsize

$$ \begin{bmatrix} e(x,y) & 0 & 0 & 0\\ 0 & - p_{y}(x,y) & 0 & 0\\  0 & 0 & - p_{z}(x,y)  & 0\\  0 & 0 & 0 & - p_{x}(x,y)\\   \end{bmatrix} =  \begin{bmatrix} \tilde{e}(x,y) & 0 & 0 & 0\\ 0 & - p(x,y) & 0 & 0\\  0 & 0 & - p(x,y)  & 0\\  0 & 0 & 0 & - p(x,y)\\ \end{bmatrix} +    \begin{bmatrix} \Pi_{00}(x,y) & 0 & 0 & 0\\ 0 &  \Pi_{11}(x,y) & 0 & 0\\  0 & 0 &  \Pi_{22}(x,y)  & 0\\  0 & 0 & 0 &  \Pi_{33}(x,y)\\ \end{bmatrix}$$

 \normalsize
\vspace{0.5cm}

The following relations are our eigenvalues which are given by

\vspace{0.7cm}
\begin{equation} e(x,y) = \tilde{e}(x,y) + \Pi_{00} \end{equation}
\begin{equation} p_y(x,y) = p(x,y) - \Pi_{11} \end{equation}
\begin{equation} p_z(x,y) = p(x,y) - \Pi_{22} \end{equation}
\begin{equation} p_x(x,y) = p(x,y) - \Pi_{33} \end{equation}

\vspace{0.7cm}

If we try to express the eigenvalues of the anisotropic stress energy-momentum tensor accordance to the Ricci components we will take

\begin{equation} e(x,y) = \tilde{e}(x,y) + \Pi_{00} = 2 \Phi_{11} + 6\Lambda + 2 \Phi_{00} \end{equation}
\begin{equation} p_y(x,y) = p(x,y) - \Pi_{11} = 2\Phi_{00} - 2\Phi_{11} - 6 \Lambda \end{equation}
\begin{equation} p_z(x,y) = p(x,y) - \Pi_{22} = 2\Phi_{11} - 6 \Lambda + 2 \Phi_{02}  \end{equation}
\begin{equation} p_x(x,y) = p(x,y) - \Pi_{33} =  2\Phi_{11} - 6 \Lambda - 2 \Phi_{02}  \end{equation}

\vspace{0.7cm}

Also, the eigenvalues take the following form which are the realtions $(55)-(58)$ accordingly.

$$ e = -d  + \frac{k_0 + l_0}{(x^2 + y^2)^3} \left[3 E^2 (y) - H^2 (x) \right]  $$

$$ p_x= d  + \frac{k_0 + l_0}{(x^2 + y^2)^3} \left[ E^2 (y) - H^2 (x) \right]  $$

$$ p_y= d  + \frac{k_0 + l_0}{(x^2 + y^2)^3} \left[- E^2 (y) - H^2 (x) \right]  $$

$$ p_z= d  + \frac{k_0 + l_0}{(x^2 + y^2)^3} \left[ E^2 (y) - 3H^2 (x) \right]  $$

After that, using the relations of $E^2 (y)$ and $H^2 (x)$ we take the completed result about the form of the eigenvalues. These forms allow us to detect the terms that are equal in relations $(56)-(58)$ of the anisotropic pressures distinguish the isotropic pressure. Hence we define as isotropic pressure the term which is equal to each of the anisotropic pressure functions. 

\vspace{1cm}
\textit{Fluid pressure}

\begin{equation} p(x,y) \equiv {d} \left[  1 + \frac{(k_0 + l_0)(y^2 - x^2)}{3 k_2  (x^2 + y^2)^2}    \right]   \end{equation} 

\vspace{1.2cm}

This term is also contained in the "energy density"\footnote{The real energy density contained in the perfect fluid energy tensor.}too. Watching closely, we see that the setting $y^2 = x^2$ reduces to a constant. Also, the same result has been obtained with the elimination of the contants $k_0, l_0$ which are related to the fluid along with the constant $d$.

\subsection{Study of the solution at the center of the star}

In this section we will see the behavior of our pressure $(99)$ at the center where the coordinates $y,x$ tend to zero. As we said before the coordinate $y$, $x$ represent the radius and the function $\alpha cos\theta$ accordingly, which are equal to zero at the center of the star since there is no angular momentum. The pressure thus goes to infinity as $x,y$ tends to zero.

This is a \textit{ring singularity} of our solution since at $y=0$ the x coordinate is equal to zero at $\theta = \pi/2$. Except the pressure, the eigenvalues goes to infinity too. These signs denote the singular profile of the solution at the center of the star. 
%%%

\subsection{Zero-Pressure surface –  The shape of our star}

The method that we used to obtain the shape of our star, is the elimination of the fluid pressure. We know that the fluid pressure must be equal to zero along the surface. We will try thus to find out in which cases the pressure is equal to zero. 

$$ p(x,y) \equiv d \left[  1 + \frac{(k_0 + l_0)(y^2 - x^2)}{3 k_2  (x^2 + y^2)^2}    \right]$$

\vspace{0.5cm}

\textit{Case 1} 

\vspace{0.5cm}
The first case concerns the elimination of the $d$ constant. This choice, as a matter of fact, coincides with the elimination of the same constant in case of the soldering in the previous chapter. Also, this condition implies different form of our functions $E^2(y)$ and $H^2(x)$ as well as our eigenvalues.

\vspace{0.7cm}
\small
\begin{equation}  e(x,y) =  \frac{(k_0+l_0)}{(x^2+y^2)^3} \left[ 3q_1 \sqrt{k_2 y^2+k_0} - q_2 \sqrt{k_2x^2+l_0} + \frac{d_2 (3y^2 +x^2)}{ k_2} -\frac{2d_0}{k_2}  +  \frac{2 d_2 (3k_0 + l_0)}{{k_2^2}} \right] \end{equation}

\begin{equation}  p_y(x,y) = - \frac{(k_0+l_0)}{(x^2+y^2)^3} \left[ q_1 \sqrt{k_2 y^2+k_0} - q_2 \sqrt{k_2x^2+l_0} + \frac{d_2 (x^2 +y^2)}{ k_2} +  \frac{2 d_2 (k_0 + l_0)}{{k_2^2}} \right] \end{equation}

\begin{equation}  p_z(x,y) =  \frac{(k_0+l_0)}{(x^2+y^2)^3} \left[ q_1 \sqrt{k_2 y^2+k_0} - 3q_2 \sqrt{k_2x^2+l_0} + \frac{d_2 (3x^2 +y^2)}{ k_2} +\frac{2 d_0}{k_2} +\frac{2d_2 (k_0 + 3l_0)}{{k_2^2}} \right] \end{equation}

\begin{equation}  p_x(x,y) =  \frac{(k_0+l_0)}{(x^2+y^2)^3} \left[ q_1 \sqrt{k_2 y^2+k_0} - q_2 \sqrt{k_2x^2+l_0} + \frac{d_2 (x^2 +y^2)}{ k_2} +    \frac{2d_2 (k_0 + l_0)}{{k_2^2}} \right] \end{equation}

\normalsize
 
If we substitute all the computed values of our contants from the previous chapter, we will take a new form for our eigenvalues. The constants were obtained are equals to the following values. 

$$l_0 = 0 \hspace{2cm} k_2 = 1$$

$$d=0 \hspace{1cm} 2d_2 = a \hspace{1cm} q_2 = b \hspace{1cm} d_0 = -c$$

\begin{equation}   y_h = -\frac{\tilde{d}}{a} \end{equation}

$$  q_1 =  - \sqrt{\tilde{d}^2 + a^2 k_0}   $$

$$c = \frac{\tilde{d}^2}{2a} $$

\vspace{1cm}

We know that the coordinate $y_h$ is a positive constant. Hence, the eigenvalues take the forms below.
\vspace{1cm}

\begin{equation}  e(x,y_h) =  \frac{k_0}{(x^2+{y_h}^2)^3} \left[    \frac{a}{2}(x^2 - {y_h}^2) - bx      \right] \end{equation}

\begin{equation}  p_y(x,y_h) = - \frac{k_0}{(x^2+{y_h}^2)^3} \left[    \frac{a}{2}(x^2 - {y_h}^2) - bx      \right] \end{equation}

\begin{equation}  p_z(x,y_h) = 3 \frac{k_0}{(x^2+{y_h}^2)^3} \left[    \frac{a}{2}(x^2 - {y_h}^2) - bx      \right] \end{equation}

\begin{equation}  p_x(x,y_h) = \frac{k_0}{(x^2+{y_h}^2)^3} \left[    \frac{a}{2}(x^2 - {y_h}^2) - bx      \right] \end{equation}

In this point, we see the existence of the eigenvalues ( pressure stresses ) at the surface. The only case, that the elimination of the quantities $p_x, p_y, p_z$ takes place surely, is the perfect-fluid case, since the hydrostatic pressure is equal to the pressure stresses. In any other case the survival of the eigenvalues is a fact. 

Also, there is a similarity between the eigenvalues at the surface. As we expect the relation $(42)$ does not been violated. 

$$p_y = -e$$ 

\begin{equation}p_z = 3 e  \end{equation}

$$ p_x = e$$

\vspace{0.5cm}

\textit{Case 2} 

\vspace{0.5cm}

The second case concerns the elimination of the term inside the brackets of the relation of the hydrostatic pressure. By this relation we can conclyde tha the shape of the surface of the star described by the Lemniscate of Bernoulli.

\begin{equation} (y^2 + x^2)^2 =2A^2 (x^2 - y^2)  \end{equation}

where we set for simplification the following equality to be equal with this positive value.

$$ \frac{k_0 + l_0}{3 k_2 } = 2A^2 $$

Now, we will use Boyer-Lindquist coordinates in order to take a more accurate look of the shape of the surface that produced by the equation (110). 

\begin{equation} (r^2 + (\alpha cos \theta)^2)^2 =2A^2 ((\alpha cos\theta)^2 - r^2)  \end{equation}

It's clear now that the surface of the pressure, where the fluid is entrapped, is described by a toroidal shape which is depended by the values of $\alpha$ and A.

\section{Discussion}

We briefly presented the main steps that were made in order to obtain the anisotropic fluid solution. The Anisotropic Fluid Solution is a global space-time model for a rotating fluid. The supplementary condition (42) is an equation of state, which produced a solvable seperated equation giving the functions $E^2 (y), H^2 (x)$. The profile of the solution is determined by the supplementary condition since another supplementary condition would change totally the result.

The term "soldering", in the seventh chapter, is referred to the matching of the exterior solution with the interior anisotropic fluid solution along the zero-pressure surface. We proved that the anisotropic fluid solution doesn't match properly with Carter's family of solutions since at the radius $y_h = - \tilde{d}/a$ the functions $E^2 (y)$ and ${E^2}_C (y)$ are equals to zero. These equalities impose that our metric component $g_{yy}$ becomes singular. Physically, this singularity implying the existence of an event horizon at the radius $-\tilde{d}/a$ where the two solutions are "matching". Maybe this singularity will be neglected, if we apply a coordinate transformation. A possible coordinate transformation from Boyer-Lindquist coordinates to Kruskal coordinates it may be proved capable to avoid the singularity. This is a possibility in which the singularity will be characterized as \textit{coordinate singularity}. Although, the validation of this hypothesis demands time.

Next, the reduction of the exterior solution to a Kerr-like solution will give us the Maximally rotating case of Kerr where the radius will be equal to mass. Hence, the result of this procedure concerns the max Kerr "stuffed" with an anisotropic rotating fluid. This is correct from one point of view since the surface of matching is the only one event horizon of max kerr. Generally, there are technics that neglects the singularity and make the matching possible via a thin shell with curious features \cite{Majidi2017}, \cite{Poisson}. This shell is used in case of existing jumps in functions that are vital for the metric like $E^2 (y)$. The existence of this thin shell, in turn, implies that $T_{\mu \nu}$ is not zero at the surface layer and the two metrics may be matched smoothly like in our eigenvalues. 

In this case, if we substitute the final values which are obtained by the matching, we observe the surviving of the eigenvalues of the stress-energy tensor (105) - (108), in conjuction with the case of perfect fluid. The non-nihilism of the eigenvalues are equivalent with the existence of discrepancies which concern unequally distributed matter, across the surface. A further investigation is needed in order to determine the nature of the matter and their physical properties as well. However, the existence of matter, in case of anisotropy, is not a prohibitive fact.

Concerning now the inner part of the solution at the last chapter, we result that there is a ring singularity where the pressure becomes singular as well as the eigenvalues of the stress-energy tensor. Also, the elimination of the pressure in the surface via the elimination of $d$ constant is a logical indication about the nature of the pressure, since the elimination of $d$ constant is the first step was made in order to obtain a Kerr-like form for our interior solution.

Also, it is generally expected that there is a variety of these kind of "stuffed" BH models, in the literature. Some of them which are contained in \cite{Brustein2017}, \cite{Hooft2016} propose to rethink about the inner of Black Holes and to revise the boundary conditions for BHs horizons. The proposal of \cite{Brustein2017} is premised on the idea that any BH-like object whose interior has some matter distribution (anisotropic generally) should support fluid modes in addition to the conventional spacetime modes. In fact, the frequency and the damping time of these modes are determined by only the object's mass and speed of rotation. These modes which are known as Coriolis-induced Rossby or r-modes are oscillated at a lower frequency and produce weaker gravitational waves than do modes of the spacetime class. Hence, they imprint an unique signature in gravitational wave spectrum of the so called "stuffed" BHs from a BH merger.

As we referred the anisotropic solution of Papakostas is an ellipsoid of revolution, this is a statement that proves the undoubtely physical profile of the solution. We know that Carter's family of solutions describes ellipsoids of revolution which have only two possible shapes depending on the observer. One family of observers can see a surface of an oblate spheroid and the other can see a toroidal. Regardless the type of the solution (interior or exterior), it is either oblate spheroid or toroidal as long as it lies in Carter's family of spaces. This theorem is contained in \cite{Papakostas2015}. This is also obvious from the second case in the last chapter. That case is referred to the case of toroidal due to its order $r^4 \propto cos^4 \theta$ .

At the end, this thesis describes the procedure of constructing a realistic model of a stuffed BH and at the end it is evinced an intesive work. The further investigation of this model, using the above suggestions in order to neglect the singularities, would be proved fruitful for various domains, like as numerical relativity. The simulation of these models (Stuffed Black Holes) would be helpful in the procedure of differentiation of their imprintations which are known as gravitational waves. The investigation of astronomical objects via their gravitational waves is promising, since could give an unique signature for every "heavy" object in the gravitational wave spectrum. Along these lines the theoretical models like ours along with numerical offerings and future observational data would be able to provide realistic models at last.

%\begin{acknowledgements}

%\end{acknowledgements}

\end{document}